\documentclass[twocolumn]{aastex631}

\begin{document}

\title{On the Superiority of the University of Arizona's Physics Club}




\author{E. Jesina}
\affiliation{Physics Club, Department of Physics, University of Arizona, Tucson, AZ, 85721}
\author{M. Harrison}
\affiliation{Physics Club, Department of Physics, University of Arizona, Tucson, AZ, 85721}
\author{H. Andras-Letanovszky}
\affiliation{Physics Club, Department of Physics, University of Arizona, Tucson, AZ, 85721}
\author{D. Krug}
\affiliation{Physics Club, Department of Physics, University of Arizona, Tucson, AZ, 85721}
\author{A. Golinkin}
\affiliation{Physics Club, Department of Physics, University of Arizona, Tucson, AZ, 85721}
\author{B. Kohn}
\affiliation{Physics Club, Department of Physics, University of Arizona, Tucson, AZ, 85721}
\author{T. Brown}
\affiliation{Physics Club, Department of Physics, University of Arizona, Tucson, AZ, 85721}
\author{A. Rose}
\affiliation{Physics Club, Department of Physics, University of Arizona, Tucson, AZ, 85721}
\author{P. Chiploonkar}
\affiliation{Physics Club, Department of Physics, University of Arizona, Tucson, AZ, 85721}


\begin{abstract}

We present an analysis on Physics Club's supremacy over the Astronomy Club at the University of Arizona. Through a thorough investigation of each club's history and content, and subsequent diligent calculations, we have proven without a doubt that the Physics Club is superior to Astronomy Club.  

\end{abstract}



\section{Introduction} \label{sec:intro}
 
The University of Arizona hosts two astrophysics-related clubs: the Astronomy Club and the Physics Club, both of which have seen extreme success in the years following the COVID-19 pandemic. However, a breakdown of the respective clubs' statistics finds a significant difference in quality between them, with Astronomy Club's quality being noticeably lower than Physics Club's. From our research into these statistics, we claim without a doubt that Physics Club is overall superior to the Astronomy Club.

We further base our accusations on previous work by \cite{barnes2002superiority} which discusses the shortcomings of Steward Observatory, the department that houses the Astronomy Club. We follow this paper as a guideline, as well as \cite{charfman2002utter}, for our approach to definitively establishing Physics Club's superiority. 

\section{Data} \label{sec:style}

\subsection{Resignations} \label{subsec: Resginations}

Over the past few years, the officer retention of Astronomy Club has been somewhat lacking. Astronomy Club has had 3 total officer resignations in the period of 2021 - 2024, whereas in the same period of time, Physics Club has had $0.5 \pm 0.5$ officer resignations. This means the rate of officer resignations per year is roughly 0.75 for Astronomy Club and only 0.125 for Physics Club. 

This shockingly high rate of resignations in Astronomy Club left the authors perplexed, as even they didn't think Astronomy Club was \textit{that} bad. However, after extensive and rigorous scientific inquiry, \footnote{Funded by Red Bull GmbH} we have determined that the officer resignation rate of Astronomy Club $\frac{dR}{dt}$ is governed by the following equation:
$$\frac{dR}{dt} = f_{baby}\frac{\omega e^{N_{Kool-Aid}}}{T}$$
where $f_{baby}$ is the percentage of club members that are pimply freshmen, $N_{Kool-Aid}$ is the amount of cults formed at star parties, $\omega$ is the lameness of presentations at meetings, and $T$ is the amount of time the officers \textit{don't} spend engaged in pointless drama. $f_{baby}$ has been observed to increase over time, while $T$ has been decreasing rapidly over the past few years, so we predict $\frac{dR}{dt}$ to increase by several orders of magnitude over the next decade. Even making the edge-case assumption that all members are willing and able to be officers, with their average attendance of 45-65 members per meeting, this officer turnover rate would quickly become unsustainable for Astronomy Club. This clearly demonstrates Physics Club's much greater potential for longevity over Astronomy Club.

\subsection{Succession} 

\begin{figure}[htb!]
    \centering
    \includegraphics[width=1.0\linewidth]{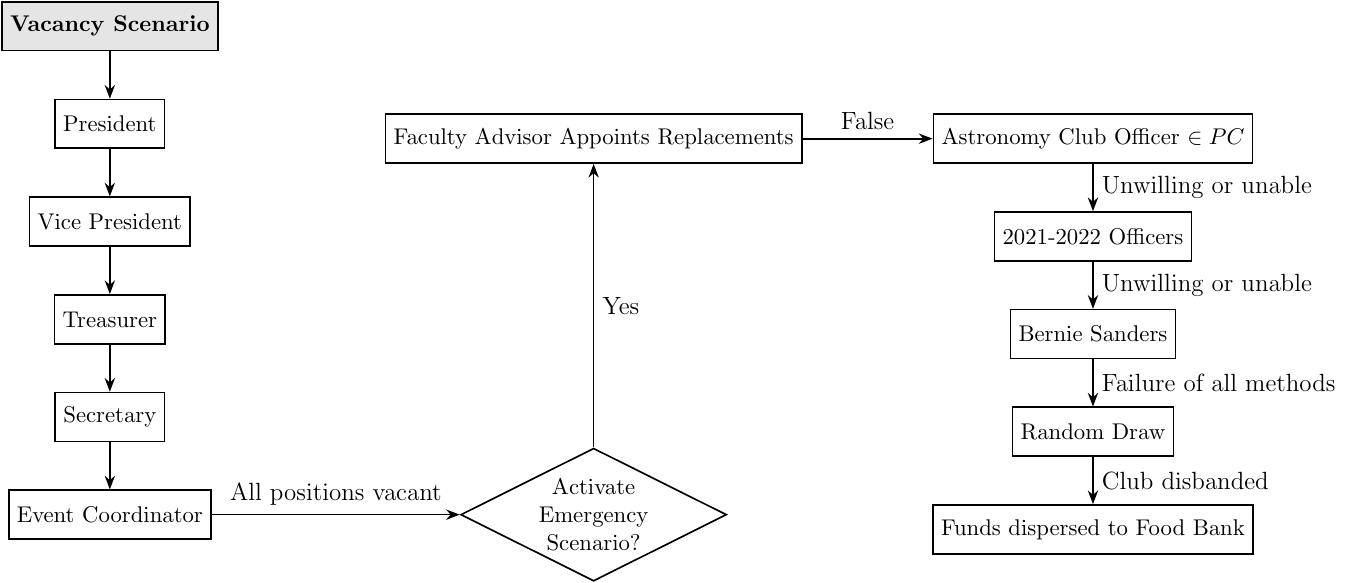}
    \caption{Flowchart of the Line of Succession}
    \label{fig:line-of-succession}
\end{figure}

The Physics Club's succession plan \citep{PhysicsClub2022} is a testament to the club's preparedness, reflecting the precision and complexity of its daily operations. While the Astronomy Club's approach to officer succession might be likened to navigating the stars without a telescope---relying on the hope that a simple special election will fill the void---the Physics Club has charted every possible eventuality and more (refer to Figure \ref{fig:line-of-succession}). Should the well-ordered cascade of leadership in cases of unexpected vacancies fail, provisions exist to secure Senator Bernie Sanders' (un)willing intervention.

Moreover, the line of succession highlights the Physics Club's spirit of generosity and cooperation. In the event of an unprecedented total leadership vacuum, the club's benevolence extends beyond the realm of academia, ensuring any remaining funds are bequeathed to the Community Food Bank of Southern Arizona---a noble act of community service rarely seen in clubs like the Astronomy Club, which often lack such foresight. The Physics Club even extends an olive branch to its Astronomy Club counterparts, allowing for an Astronomy officer (who is also a Physics Club member) to take the helm, should the need arise. This gesture of cross-club solidarity, however, appears to be a one-way street, as no such reciprocity is found in the Astronomy Club's statutes \citep{AstronomyClub2023}.

\subsection{Memes}

It has been known since time immemorial that the quality and quantity of memes held within the Astronomy Club's slides have been below that of the Physics Club, but there have been no attempts to compare these memes until now. To go about our quantitative comparison, we went through the past year's worth of available Astronomy and Physics Club slides, tallying the total amount of memes in both.

\begin{figure}[htb!]
    \centering
    \includegraphics[width=1\linewidth]{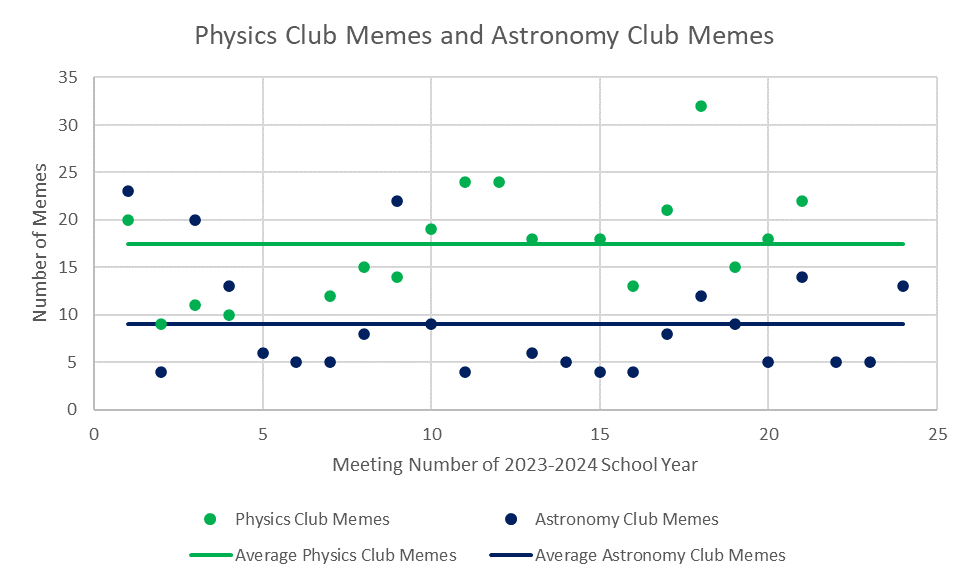}
    \caption{Number of Memes per Meeting for both Astronomy Club and Physics Club and Average Number of Memes for both Astronomy Club and Physics Club. While there have certainly been occasions where the Astronomy Club memes per meeting have surpassed the Physics Club average, this is a rarity and should be discounted as an anomaly.}
    \label{fig:meme_count}
\end{figure}
We counted a total of 209 memes for Astronomy Club during the period between the 25th of August 2023 and the 29nd of March 2024, having a total of 23 meetings during that time, with an average of $9.087$ memes per meeting. In comparison, during that same period, we counted 315 memes for Physics Club while having only 18 meetings, with an average of $17.5$ memes per meeting, almost twice the number of average Astronomy Club memes (see Figure \ref{fig:meme_count}). Thus, the sheer domination the Physics Club has over the Astronomy Club when it comes to the quantity of memes was clearly obvious to the authors.





\subsection{Demos}

Physics Club has a known affinity for demonstrations (lovingly referred to as "demos"), which has led to merry destruction on many occasions. Most notable is the Pumpkin Smashing, an annual tradition where pumpkins are frozen in liquid nitrogen and then flung from high places, such as ladders. Another common phenomenon is fire demos, where Physics Club members light objects on fire in an increasingly entertaining number of ways, including rockets, arson, melting coins, and glassblowing. As mentioned previously, the club makes frequent use of liquid nitrogen, which has been used to make ice cream, shatter assorted fruits and vegetables, and even make superconductors.

\begin{figure}[ht!]
    \centering
    \includegraphics[width=1\linewidth]{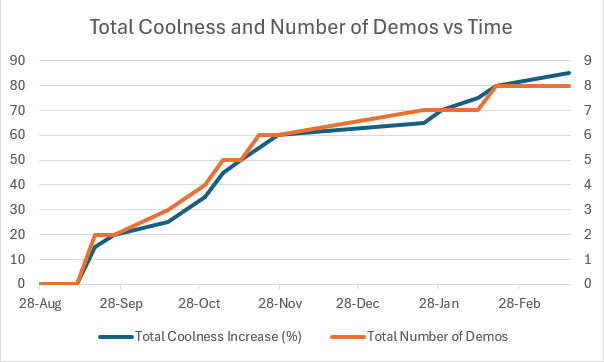}
    \caption{Total increase in coolness of Physics Club and total number of demos over the 2023-2024 school year. There is a clear direct correlation of the amount of liquid nitrogen spilled on the floor of PAS 201 and the overall coolness of Physics Club's members.}
    \label{fig:demo_coolness}
\end{figure}

It is obvious that Physics Club has great success with attracting attention with demos. However, in astronomy, most celestial objects are very far away. This makes it hard to do demos. The one thing Astronomy Club does do is something called "stargazing", an archaic act where they look at objects very far away that have absolutely no effect on anybody on Earth. This is very boring, as these objects cannot be lit on fire or doused in liquid nitrogen, meaning that there is no point to this "stargazing". Physics Club is able to do many demos, and this makes us even cooler than liquid nitrogen (see Figure \ref{fig:demo_coolness}) and hotter than fire. This lack of allure is why Astronomy Club cannot attract new members or impress their current ones.

\subsection{Professor Talks}

The superiority of Physics Club can also be seen through the amount of Notable Interdisciplinary Profound Professorial Lectures Exemplifying Science (NIPPLES) it has held.  In the past 63,113,472 seconds alone, Physics Club has had 14 NIPPLES, whereas Astronomy Club's NIPPLES count lies at a measly $1 \pm 0.2$. The discrepancy in these values suggests that the members of Physics Club are much more learned than those of Astronomy Club.

Another thing of note in the NIPPLES of Physics Club is the vast swaths of fields that these lectures have touched upon.  Some notable NIPPLES have integrated physics with global warming, computer science, and biology.  The singular NIPPLES topic of Astronomy Club in the past 63,113,472 seconds was on the subject of space rocks.

These findings indicate that the members of Physics Club, with their vast knowledge of multiple topics, are akin to the Renaissance polymaths of the past; those of Astronomy Club could be likened to a medieval farmer whose brain consists of nothing more than $70$ kB of farming knowledge.

\section{Conclusion}
We have definitively shown that the University of Arizona's Physics Club is superior over the Astronomy Club. This is obvious through the lack of resignations, the dedication to the community and emergency preparedness, the quantity of memes, amount of fire in demonstrations, and overall scientific relevance. 

\begin{acknowledgments}

The University of Arizona's Physics Club would like to thank the Astronomy Club for their promotion, generosity, and sense of humor over the years. We would also like to thank the former officers for paving the way for the current clubs to be successful. 

\end{acknowledgments}

\software{
\LaTeX, Python, PGF/TikZ, Microsoft Paint, Microsoft Excel.
}

%

\vspace{5mm}





\bibliography{sample631}{}
\bibliographystyle{aasjournal}



\end{document}